\documentclass[a4paper]{article}

\usepackage{icrc2013}

\usepackage[english]{babel}

\title{Full Monte Carlo simulations of radio emission from extensive air showers with CoREAS}

\shorttitle{T.\ Huege, C.W.\ James --- Simulating radio emission from EAS with CoREAS}

\authors{T.~Huege$^{1}$, C.W.~James$^{2}$}
\afiliations{$^1$IKP, Karlsruhe Institute of Technology (KIT), Postfach 3640, 76021 Karlsruhe, Germany\\
             $^2$ECAP, University of Erlangen-N\"urnberg, Erwin-Rommel-Str. 1, 91058 Erlangen, Germany}
\email{tim.huege@kit.edu}

\abstract{CoREAS is a Monte Carlo simulation code for the calculation of radio emission from extensive air showers. It is based on the ``endpoint formalism'' for radiation from moving charges implemented directly in CORSIKA. Consequently, the full complexity of the air-shower physics is taken into account without the need for approximations or assumptions on the emission mechanism. We present results of simulations for an unthinned shower performed with CoREAS for both MHz and GHz frequencies. At MHz frequencies, the simulations predict the well-known mixture of geomagnetic and charge excess radiation. At GHz frequencies, the emission is strongly influenced by Cherenkov effects arising from the varying refractive index in the atmosphere. In addition, a qualitative difference in the symmetry of the GHz radiation pattern is observed when compared to the ones at lower frequencies. We also discuss the strong increase in the ground area subtended by the radio emission when going from near-vertical to very inclined geometries, making very inclined air showers the most promising ones for cosmic ray radio detection.}


\begin{document}
\maketitle

\section{Introduction}

The modelling of radio emission from extensive air showers has made great progress in the past few years. With CoREAS 1.0 \cite{HuegeARENA2012a}, we have developed a Monte Carlo simulation code which takes into account the full complexity of cosmic ray air-shower physics and predicts the radio emission from charged particles in the cascade on the basis of pure classical electrodynamics. No specific ``radio emission mechanism'' has to be assumed, and the simulation is free of any tunable parameters. Therefore, CoREAS can be used for quantitative predictions of air shower radio emission which can be compared directly with experimental results \cite{SchroederIcrc2013AERA,SchroederIcrc2013LOPES}.

CoREAS makes use of the ``endpoint formalism'' \cite{JamesFalckeHuege2010} for the calculation of the electromagnetic radiation directly in CORSIKA \cite{HeckKnappCapdevielle1998}. In this formalism, any acceleration of charged particles leads to radiation which is then superposed for all particles in the cascade. The effects of the refractive index gradient in the atmosphere are correctly taken into account and have interesting consequences for the emission in particular at frequencies above 100~MHz. A wealth of options provided by CORSIKA, such as the different available hadronic interaction models, the configurable atmospheric profile, and the curved geometry of the atmosphere, are fully supported.

In the following, we present some results gathered with the CoREAS code. We first show the emission predicted by CoREAS for an \emph{unthinned} air shower at the site of the LOPES \cite{FalckeNature2005} and CROME \cite{SmidaIcrc2013CROME} experiments. Afterwards, we discuss how strongly the zenith angle of the extensive air shower influences the size of the radio emission footprint, making very inclined air showers the most interesting target for large-scale air-shower radio detection.

\section{Simulations for an unthinned shower}

We have simulated a vertical air shower with an energy of $10^{17}$~eV induced by an iron primary at the site of the LOPES and CROME experiments (110~m above sea-level, geomagnetic field of 48~nT with 65$^{\circ}$ inclination). 320 antenna locations positioned on concentric rings with a radial step size of 5 metres were simulated. The interaction models used in CORSIKA were QGSJETII.03 and UrQMD1.3.1. The shower was simulated entirely without particle thinning --- the presented simulation is thus truly a ``full Monte Carlo simulation'' of the radio emission of an extensive air shower.

In figure \ref{fig:mapsvertical}, we present maps of various components of the electric field vector after filtering to observing frequency windows used by various experiments. From top to bottom, maps for the projections of the electric field vector on the north-south, east-west and vertical axes are shown, followed by the absolute amplitude of the complete vector. From left to right, the observing frequency windows of 40 to 80~MHz (LOPES \cite{FalckeNature2005}), 300 to 1200~MHz (ANITA \cite{HooverNamGorham2010}) and 3.4 to 4.2~GHz (CROME \cite{SmidaIcrc2013CROME}) are shown. To create these maps, the raw output of CoREAS, corresponding to time-pulses with an unlimited observing bandwidth, were digitally filtered to the desired frequency windows. Afterwards, the maximum amplitude of the given electric field vector component was read-off and normalized by the effective bandwidth.

\subsection{Emission at tens of MHz}

Looking at the first column of figure \ref{fig:mapsvertical}, the typical emission characteristics as observed at MHz frequencies become apparent. The signal is consistent with a superposition of Askaryan charge-excess radiation \cite{Askaryan1962a,Askaryan1965} and geomagnetic radiation \cite{KahnLerche1966}, as is by now also evident from experimental data \cite{HuegeIcrc2013AERA}. The former contribution has linear polarization with an electric field vector oriented radially with respect to the shower axis. The latter has linear polarization with an electric field vector aligned in the direction given by the Lorentz force, i.e., in the east-west direction for a vertical air shower. In the case discussed here, the north-south component of the electric field is thus generated purely by Askaryan emission, whereas the east-west component constitutes a superposition of the two components, leading to the well-known east-west asymmetry in the signal. The vertical component of the electric field is very small, but not exactly zero, particularly very close to the shower axis. The predicted field strengths of $\approx$ 2 $\mu$V/m/MHz approximately correspond to the detection threshold of the LOPES experiment in the noisy environment of the Karlsruhe Institute of Technology \cite{SchroederIcrc2013LOPES}.

\subsection{Emission at hundreds of MHz}

In the second column of figure \ref{fig:mapsvertical}, the results for a frequency window of 300 to 1200 MHz are shown. This is the frequency band of the ANITA experiment, which has reported the successful detection of several cosmic ray events in its second flight \cite{HooverNamGorham2010}. Polarization-wise, the same superposition of Askaryan charge-excess and geomagnetic emission is apparent. However, the emission pattern is strongly influenced by Cherenkov effects arising from the density gradient and thus varying refractive index of the atmosphere \cite{AllanRefractive1971,DeVriesBergScholten2011}. This refractive index gradient changes the coherence conditions of the radiation. For an observer on the ``Cherenkov ring'' visible in the maps, the time-pulses of the Askaryan and geomagnetic radio emission are compressed to very short time-scales, leading to significant power at frequencies as high as hundreds of MHz. \footnote{It should be stressed that the Askaryan effect is not ``classical'' Cherenkov radiation in the sense of unaccelerated particles propagating through a medium with a velocity higher than the speed of light in the medium \cite{JamesFalckeHuege2010}. In fact, ``classical'' Cherenkov radiation contributes negligibly to the radio-emission from particle cascades and is not modelled at all in CoREAS and comparable models.} The diameter of the ring is governed by the geometrical distance of the shower maximum to the observer, and is thus directly related to the depth of the air shower shower maximum $X_{\mathrm{max}}$. The spectral field strength drops about an order of magnitude with respect to the values at tens of MHz. To judge the detectability, this drop has to be compared with the evolution of Galactic and atmospheric noise as a function of rising frequency. The vertical electric field component is non-zero and exhibits both a Cherenkov ring and an east-west asymmetry.

\subsection{Emission at a few GHz}

\begin{figure*}[!htb]
  \vspace{2mm}
  \centering
  \includegraphics[width=0.32\textwidth]{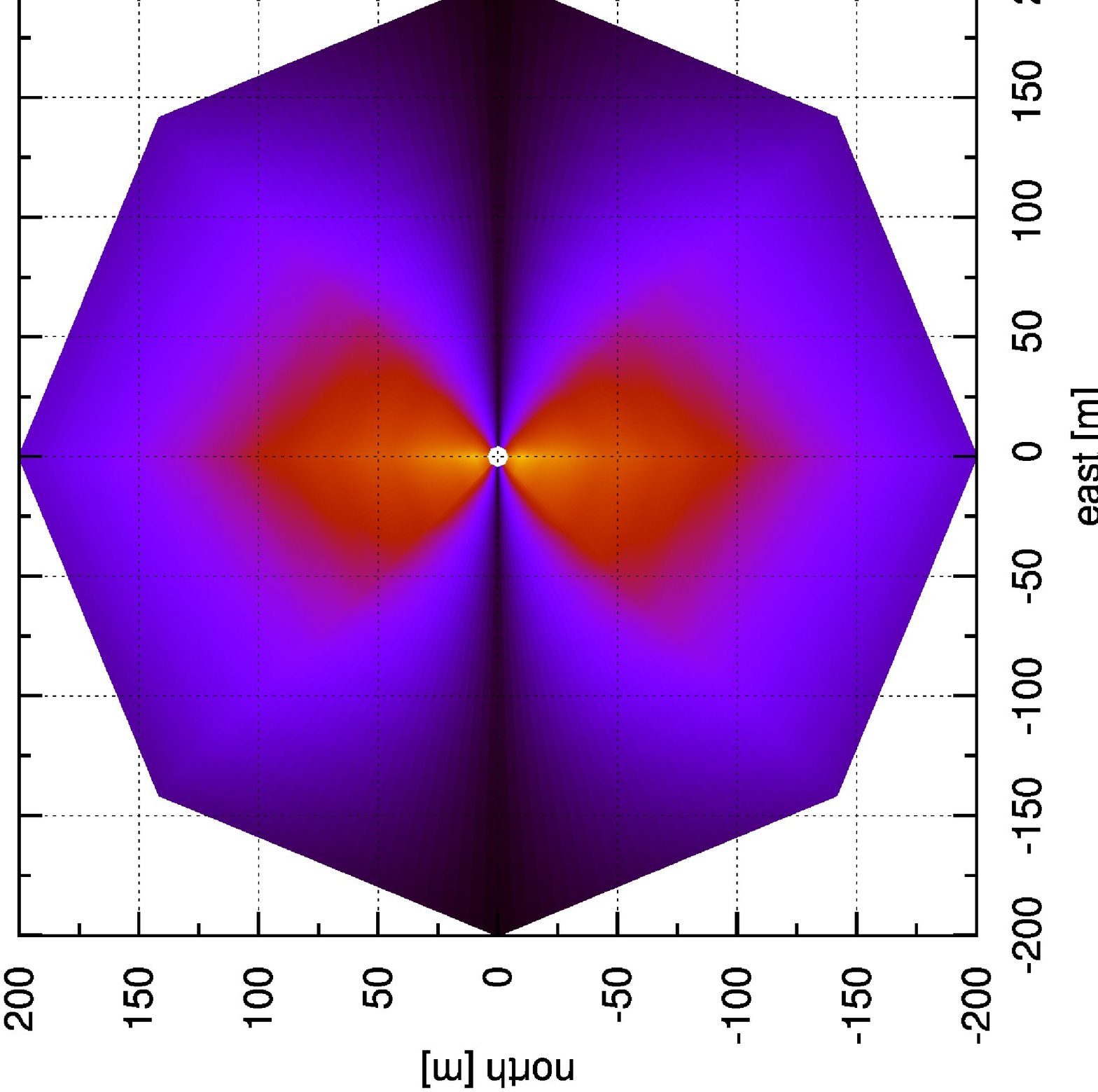}
  \includegraphics[width=0.32\textwidth]{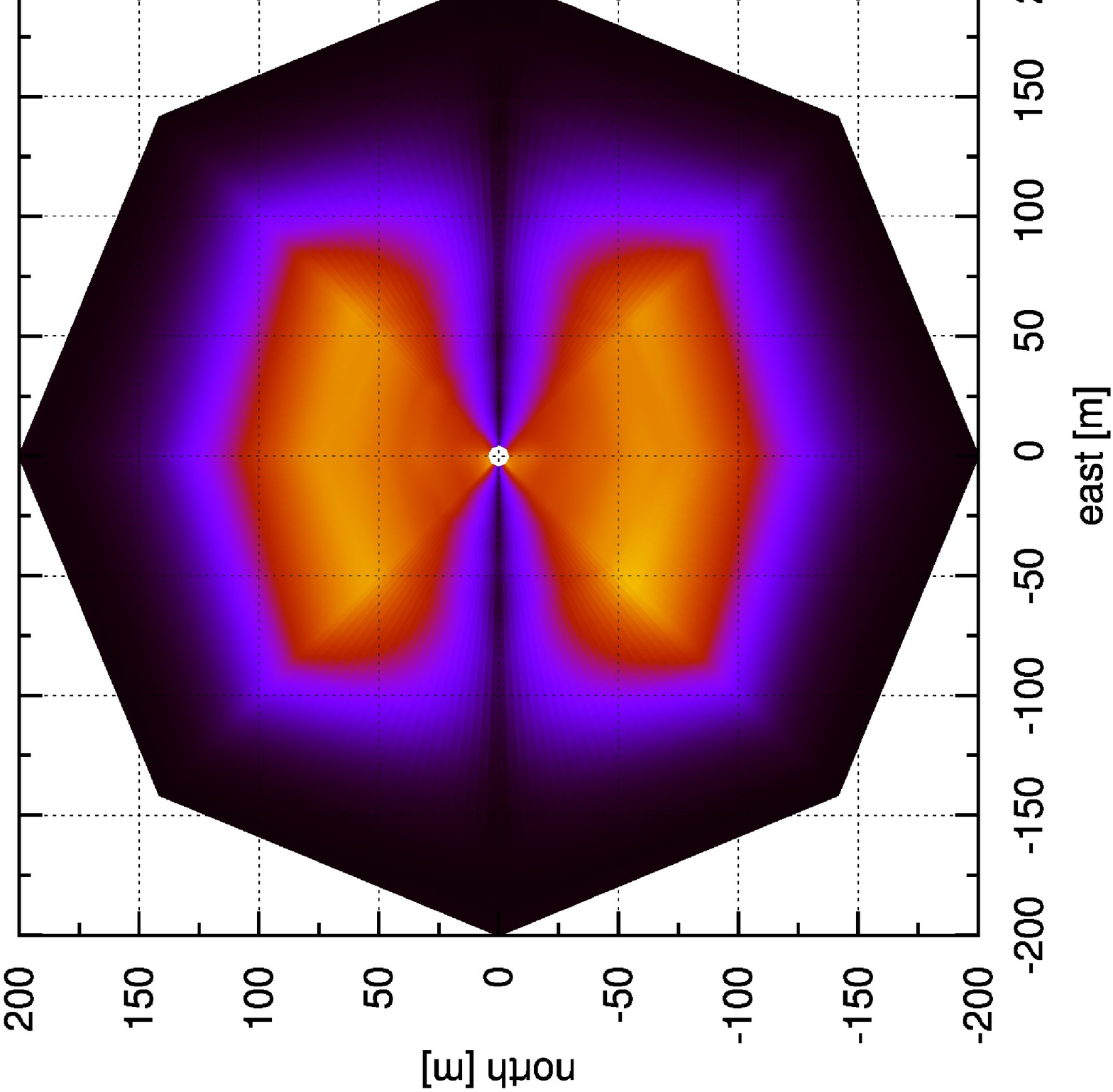}
  \includegraphics[width=0.32\textwidth]{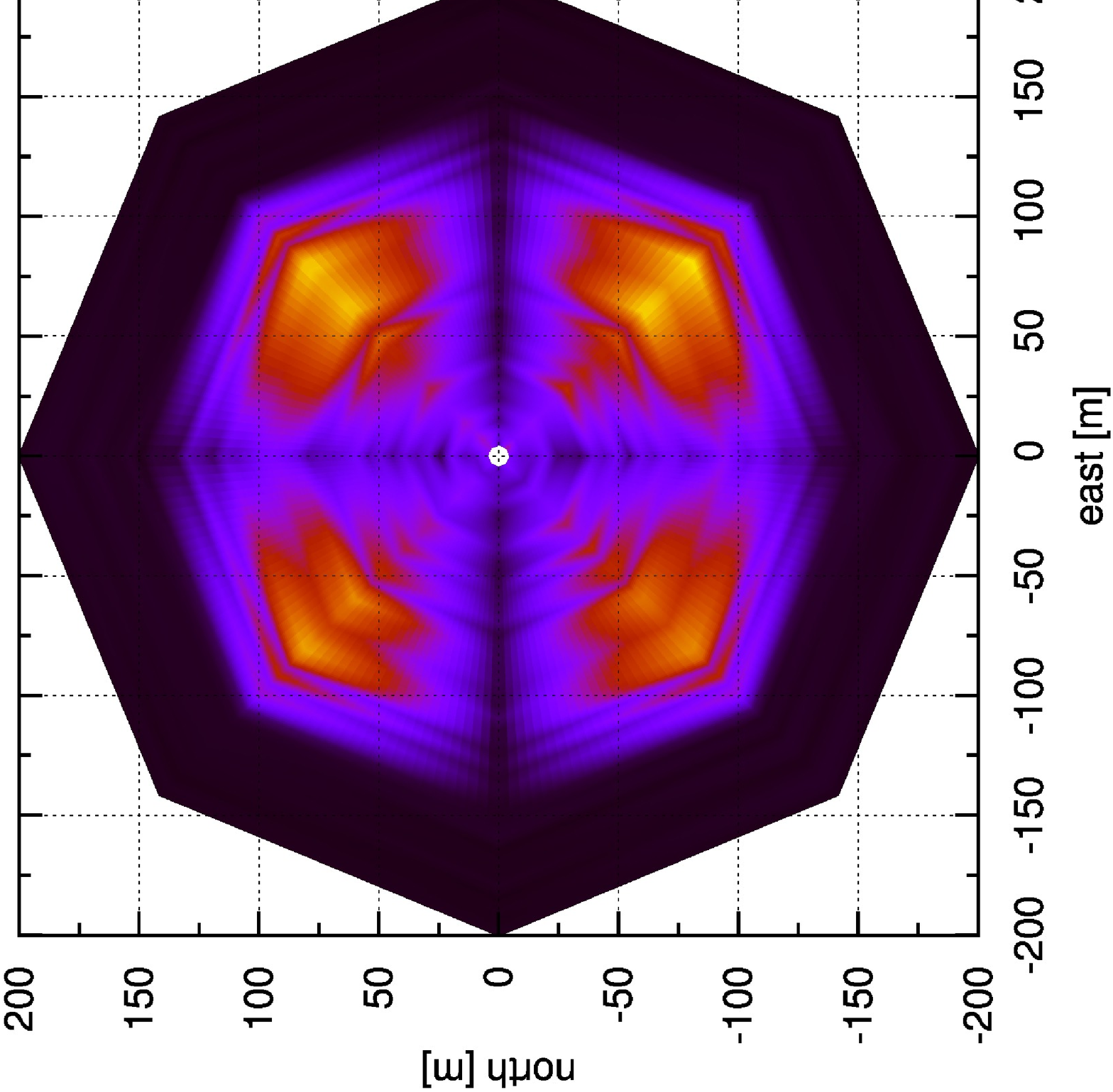} \\
  \includegraphics[width=0.32\textwidth]{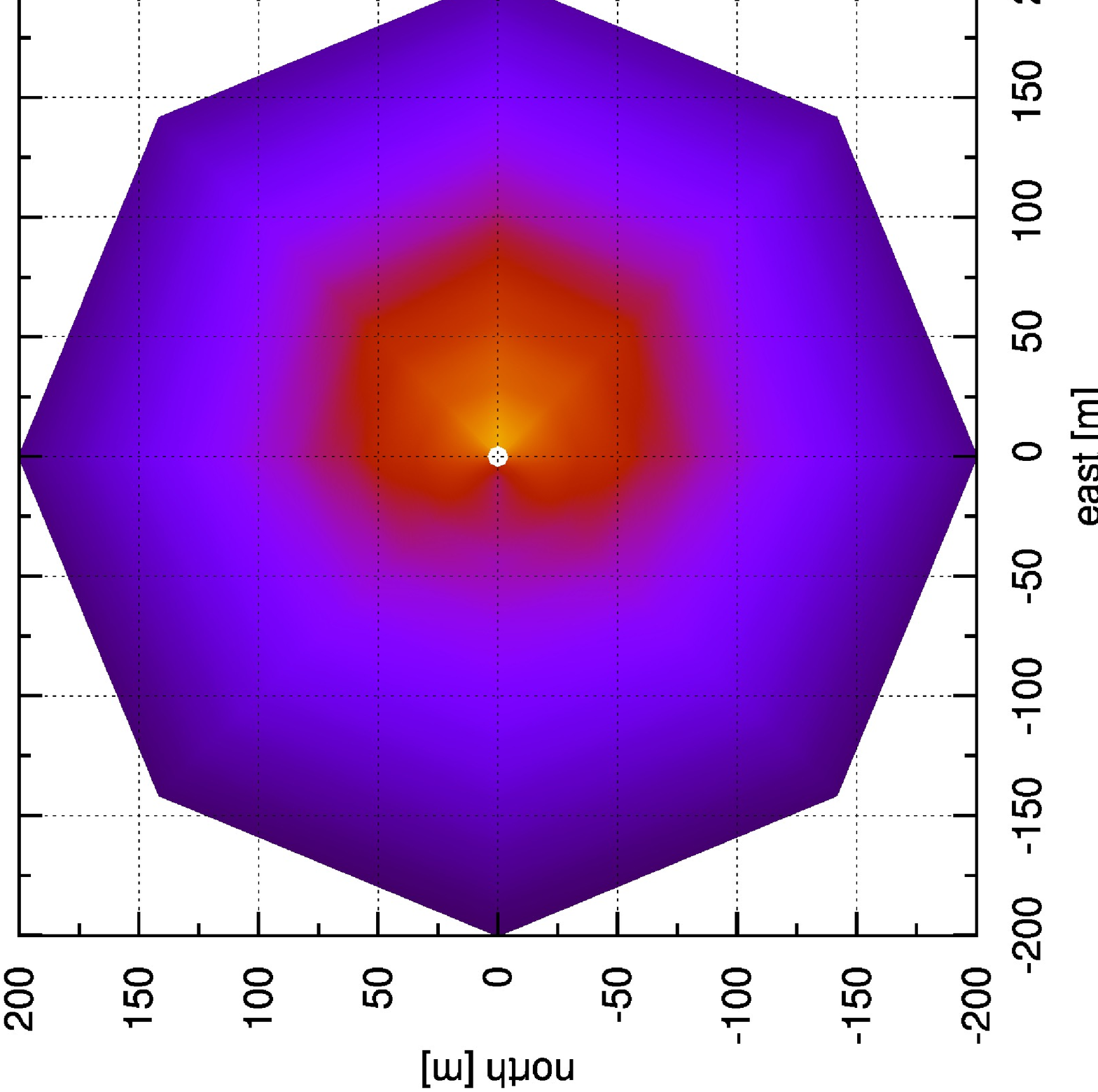}
  \includegraphics[width=0.32\textwidth]{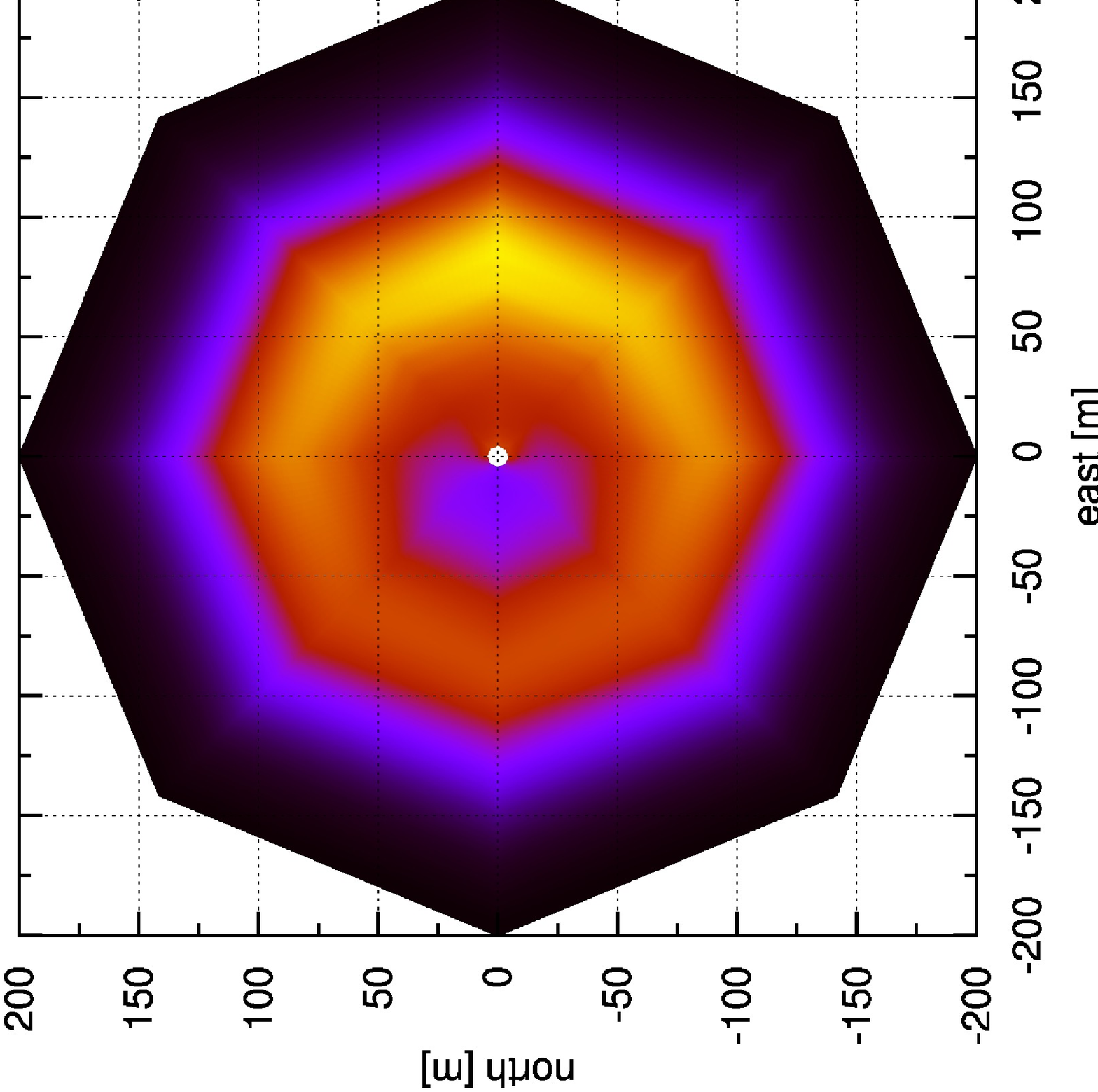}
  \includegraphics[width=0.32\textwidth]{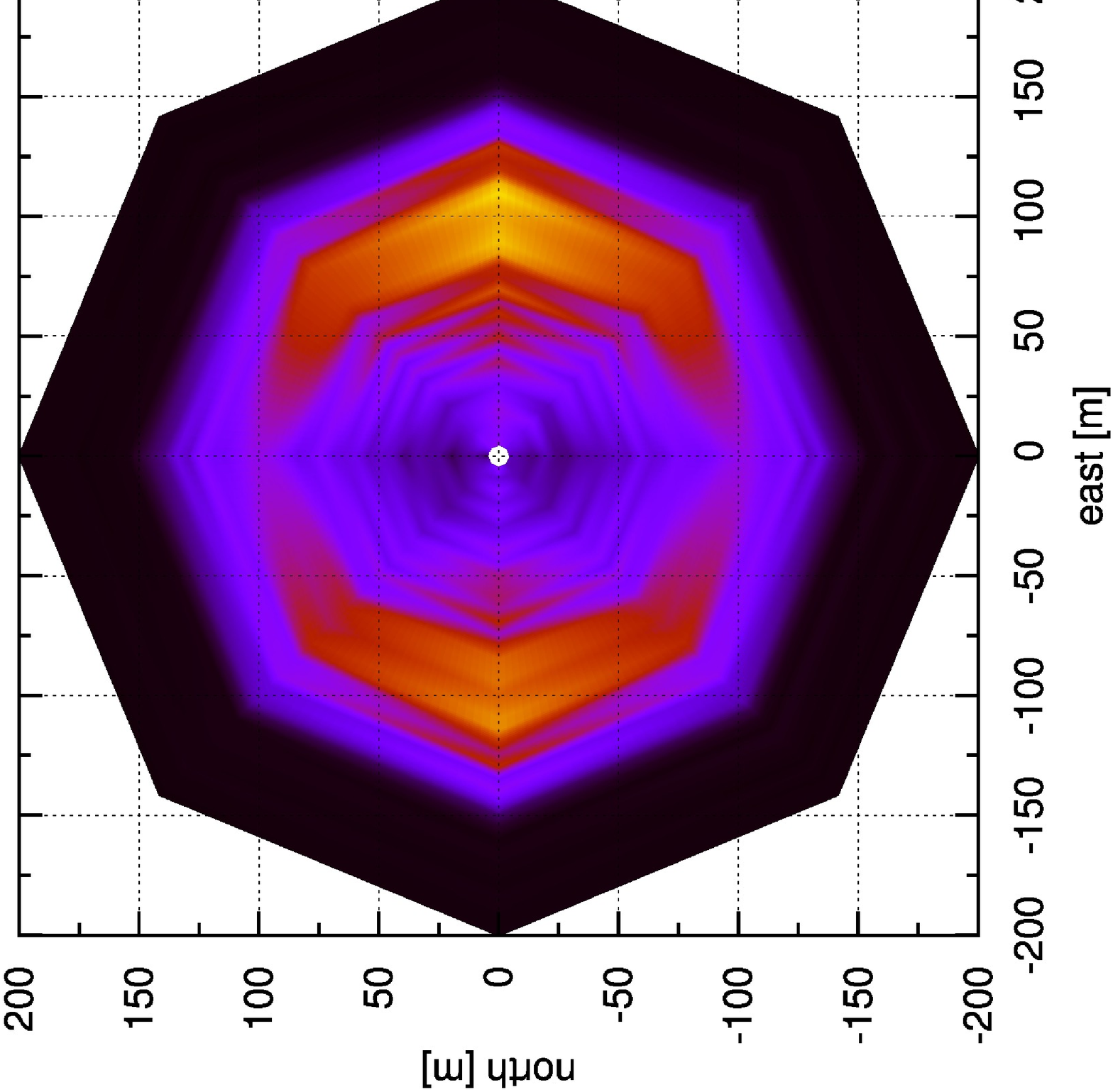} \\
  \includegraphics[width=0.32\textwidth]{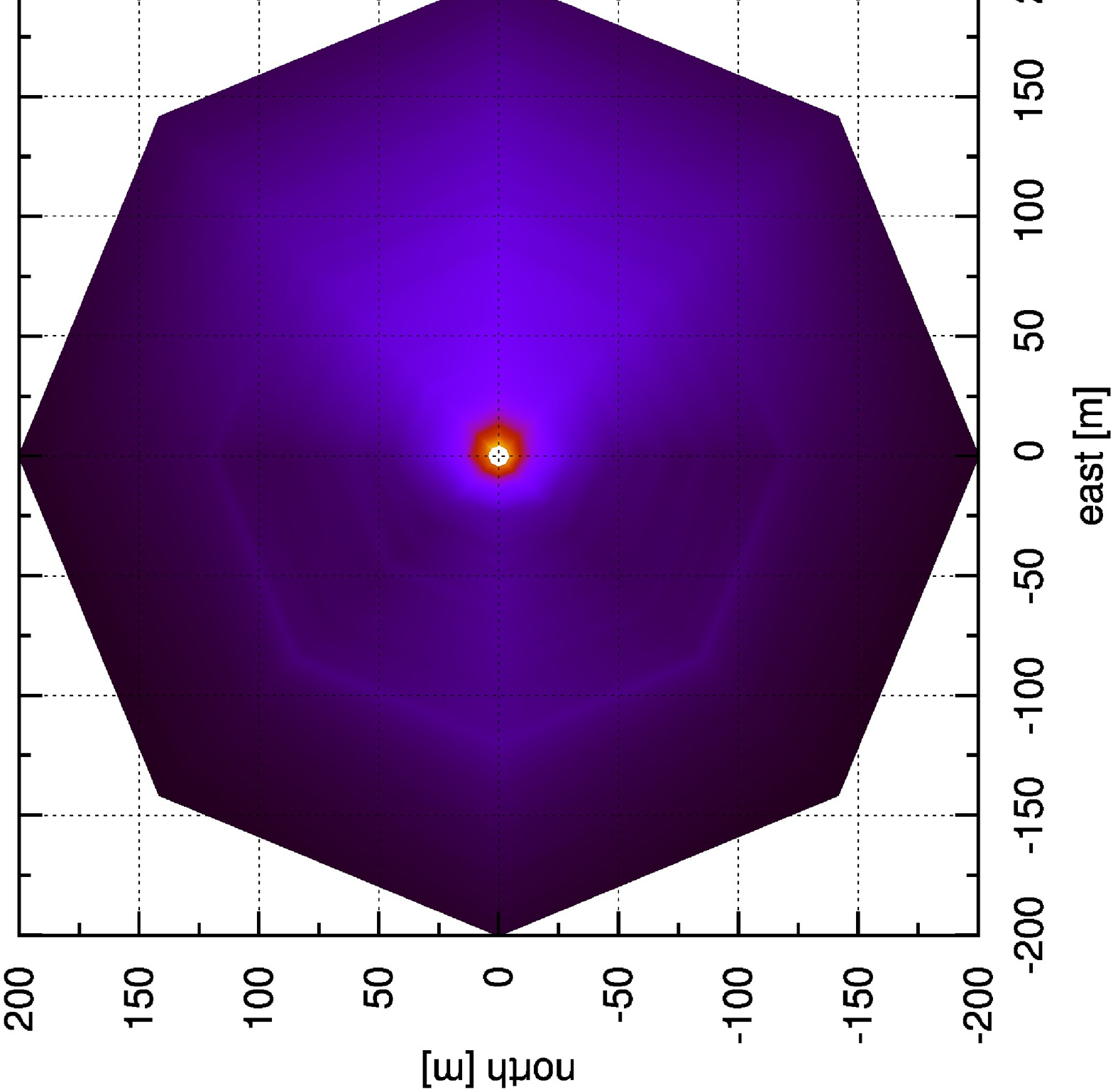}
  \includegraphics[width=0.32\textwidth]{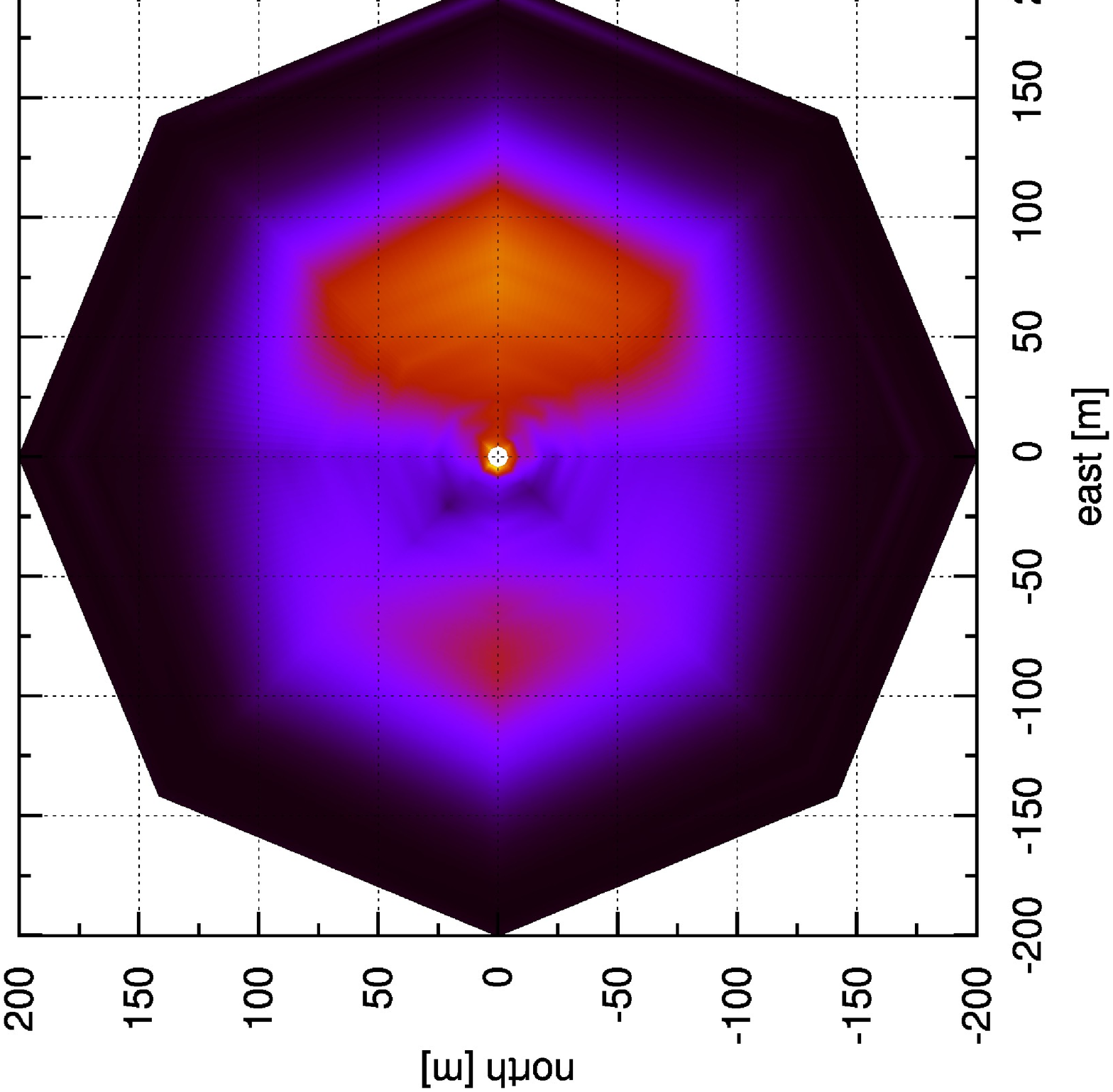}
  \includegraphics[width=0.32\textwidth]{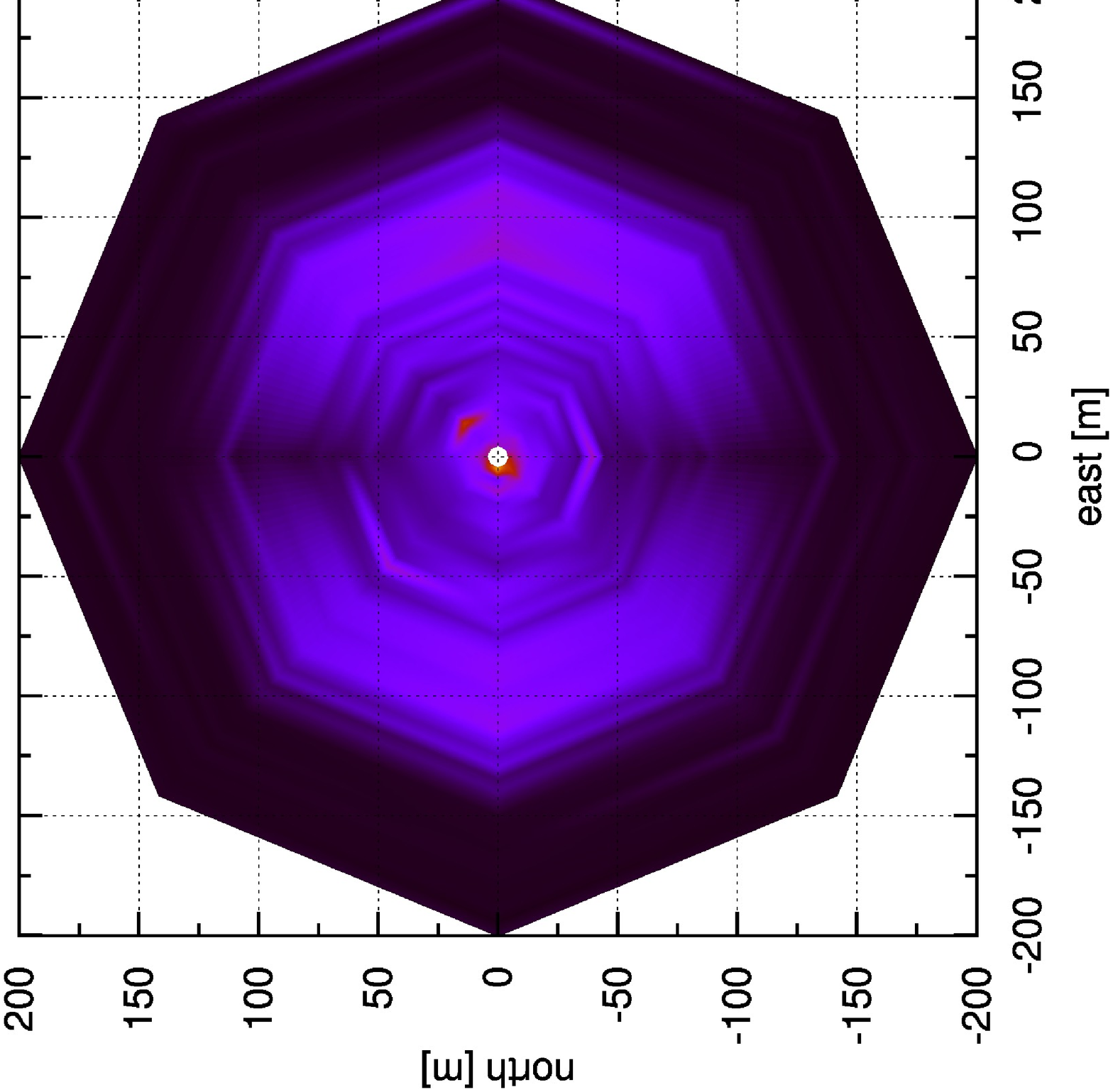}\\
  \includegraphics[width=0.32\textwidth]{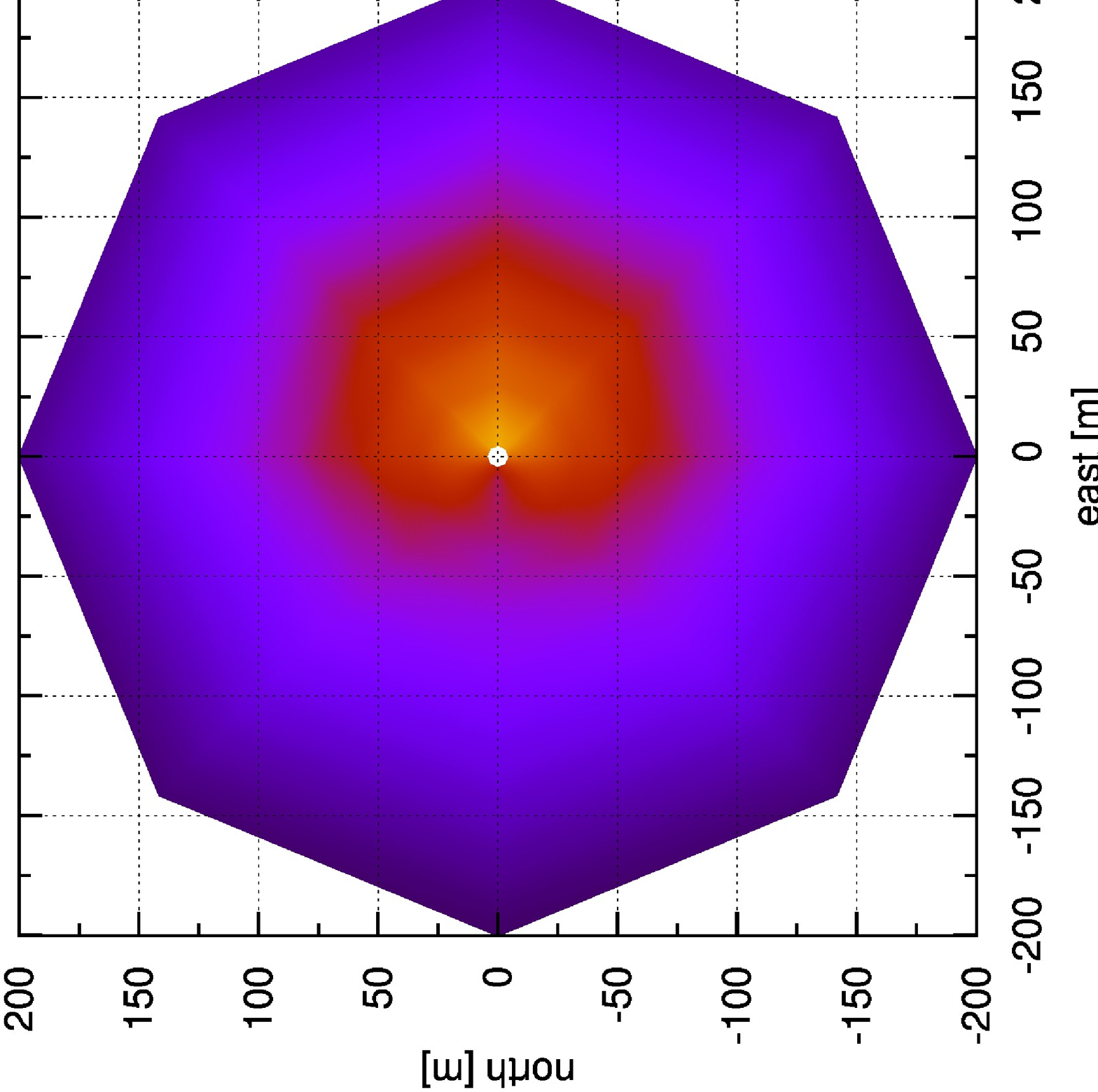}
  \includegraphics[width=0.32\textwidth]{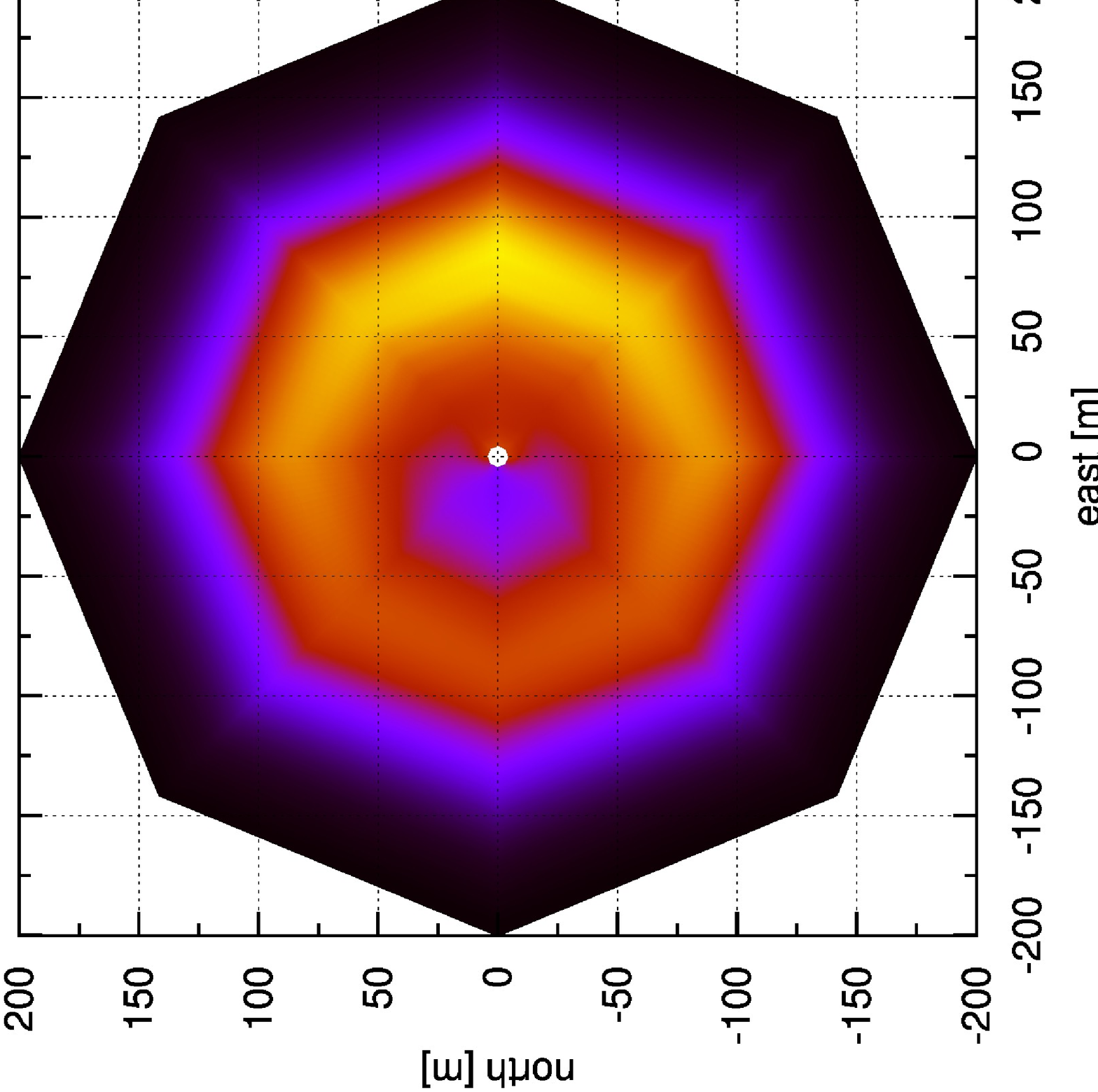}
  \includegraphics[width=0.32\textwidth]{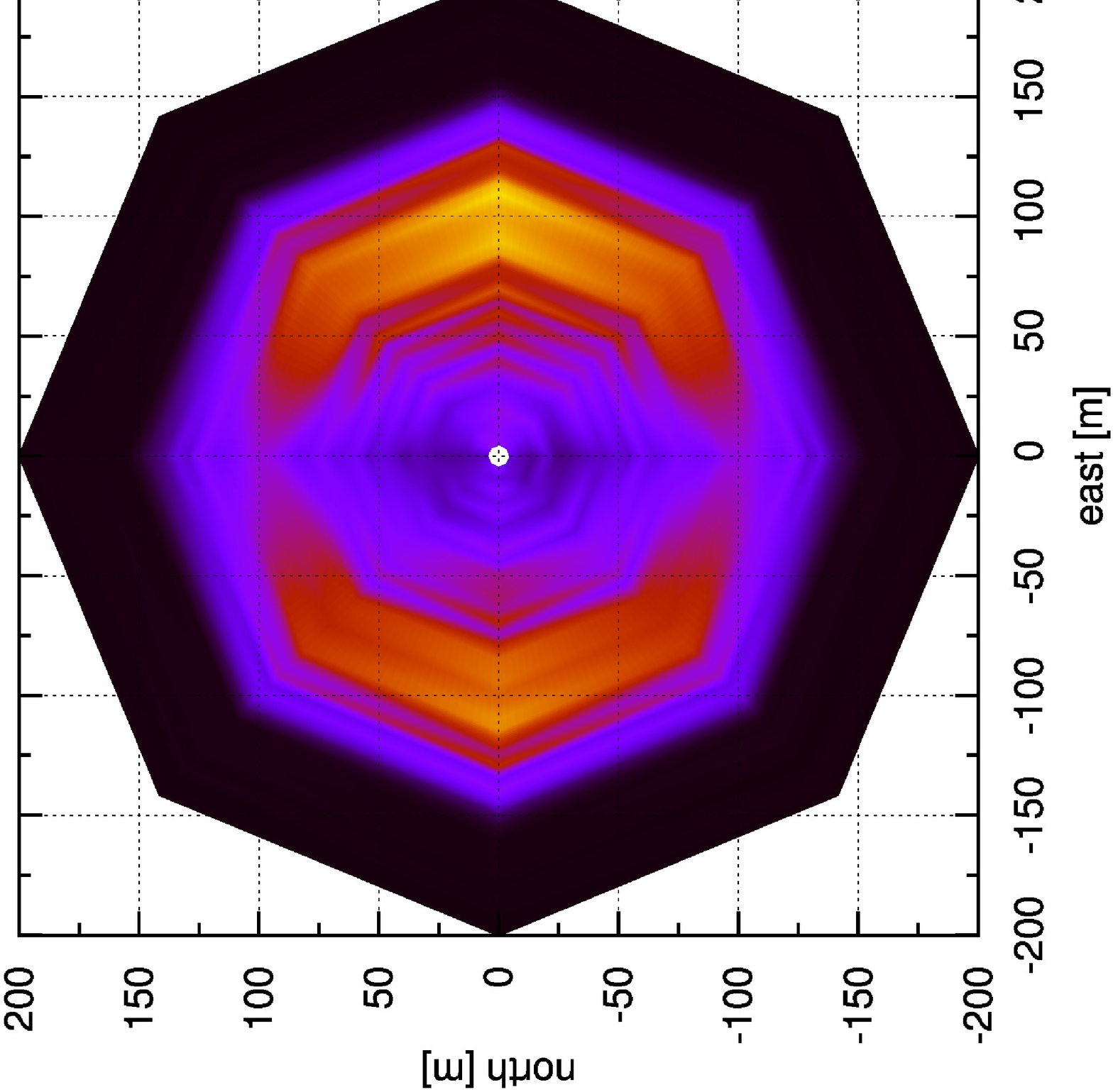}
  \caption{Maps for the spectral field strength of the maximum radio pulse plotted for different components of the electric field vector and after filtering to various observing frequency windows. From top to bottom: north-south component, east-west component, vertical component, total electric field amplitude. From left to right: 40 to 80 MHz, 300 to 1200 MHz and 3.4 to 4.2 GHz. In the white regions at the center, no simulation was carried out.}
  \label{fig:mapsvertical}
 \end{figure*}

In the third column of figure \ref{fig:mapsvertical}, we present the signal predicted for a frequency window of 3.4 to 4.2~Ghz. This is the frequency range probed by the CROME experiment and other experiments originally designed to search for ``Molecular Bremsstrahlung'' radiation at GHz frequencies. Due to the same Cherenkov-effects compressing the Askaryan and geomagnetic emission to short time-scales, significant signal levels can be visible for observers sitting on the ``Cherenkov ring'', although the spectral amplitude is again significantly smaller than at hundreds of MHz. In fact, CROME has detected such events \cite{SmidaIcrc2013CROME} and made comparisons with CoREAS simulations \cite{WernerIcrc2013CROME}. It should be noted that the CoREAS simulation code was developed primarily with the MHz regime in mind, but it clearly has predictive power also at these much higher frequencies.

While the emission at GHz frequencies is still dominated by the geomagnetic emission with the Askaryan charge-excess emission being a secondary effect (as has been verified with a simulation without magnetic field, not shown here), there are notable differences in the emission pattern as compared to lower frequencies. The Cherenkov ring is visible in the dominant east-west component of the signal, but the ring is ``broken'' along the north-south observer axis. Even more interestingly, the north-south component of the radio emission shows a ``clover-leaf'' pattern \cite{HuegeARENA2012a}. When the geomagnetic field is switched off, the clover-leaf pattern vanishes. This means that the north-south component has a contribution due to geomagnetic emission, unlike the patterns seen at lower frequencies, where the geomagnetic emission contributes purely to the east-west component of the electric field. Such a clover-leaf pattern in the north-south component of the electric field was predicted by the early ``geosynchrotron emission'' models \cite{HuegeFalcke2005b,SuprunGorhamRosner2003}. A possible explanation is that the ``geosynchrotron effect'' is significant at high frequencies only, whereas it is swamped by the emission from time-varying transverse currents at lower frequencies. Other interesting features are the ``ripples'' visible in the emission pattern. As the shower is unthinned, these appear to be interference effects of some sort.

\section{Zenith-angle dependence}

\begin{figure*}[!htb]
  \vspace{2mm}
  \centering
  \includegraphics[width=0.31\textwidth]{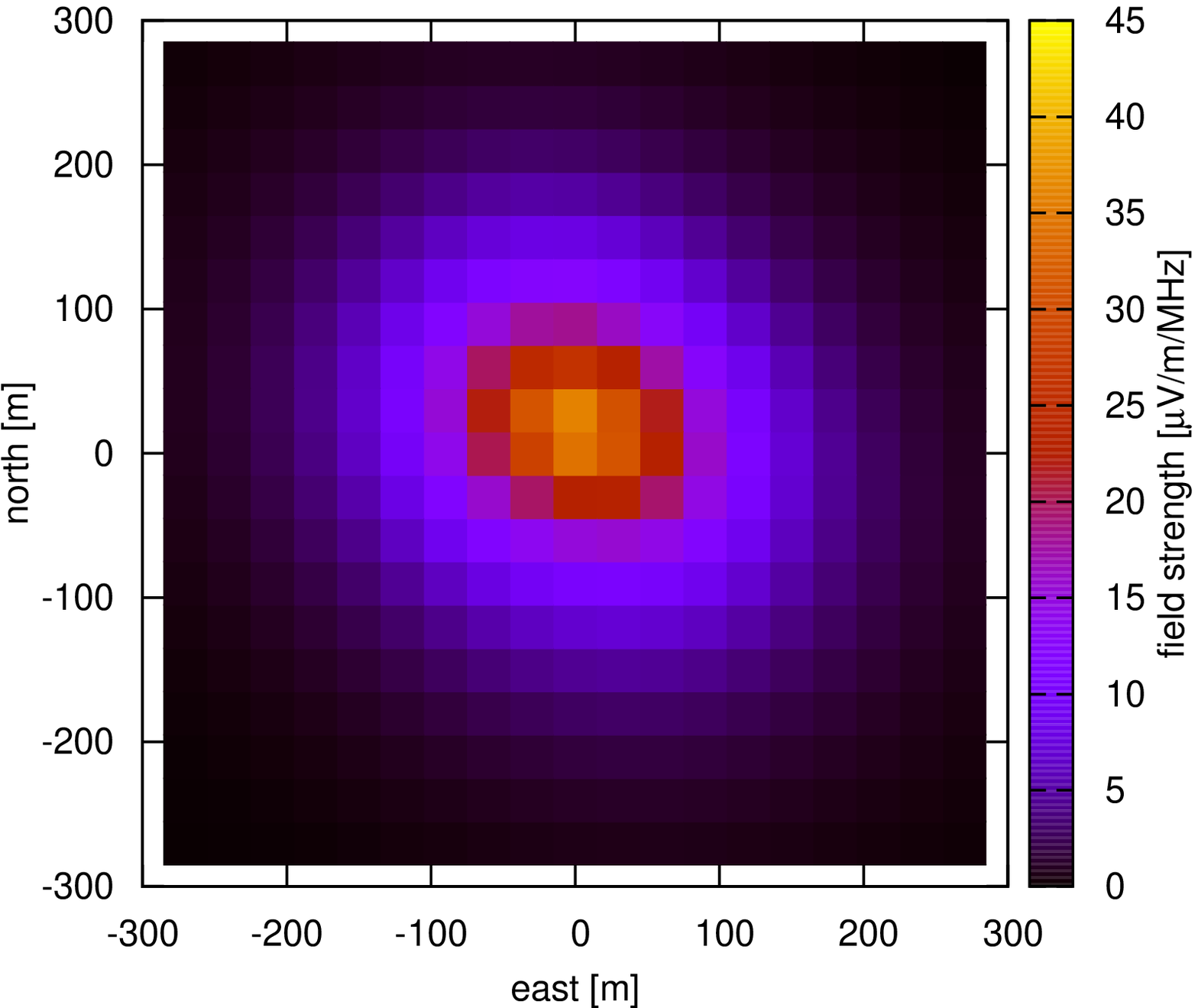} \hspace{0.2cm}
  \includegraphics[width=0.31\textwidth]{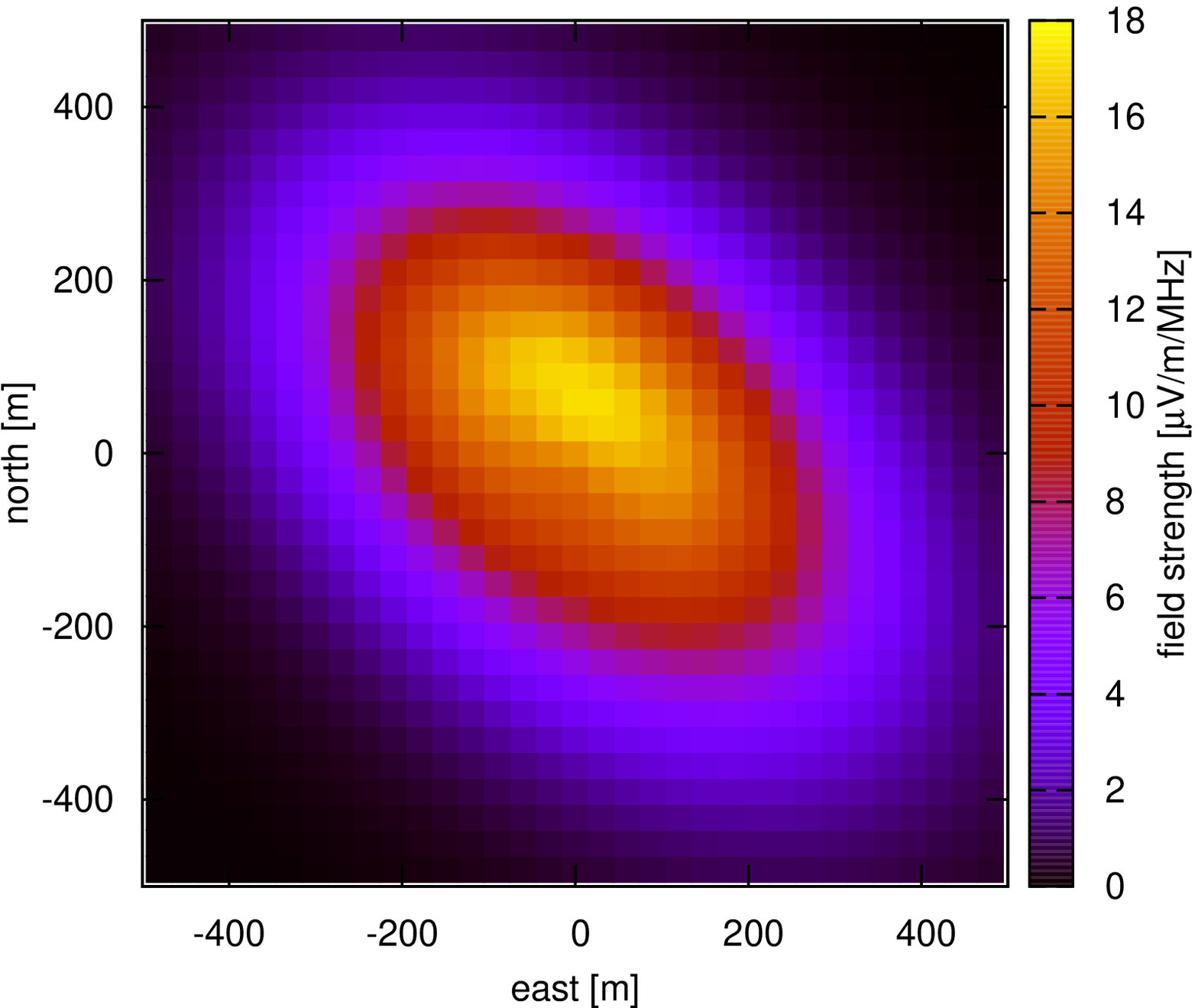} \hspace{0.2cm}
  \includegraphics[width=0.31\textwidth]{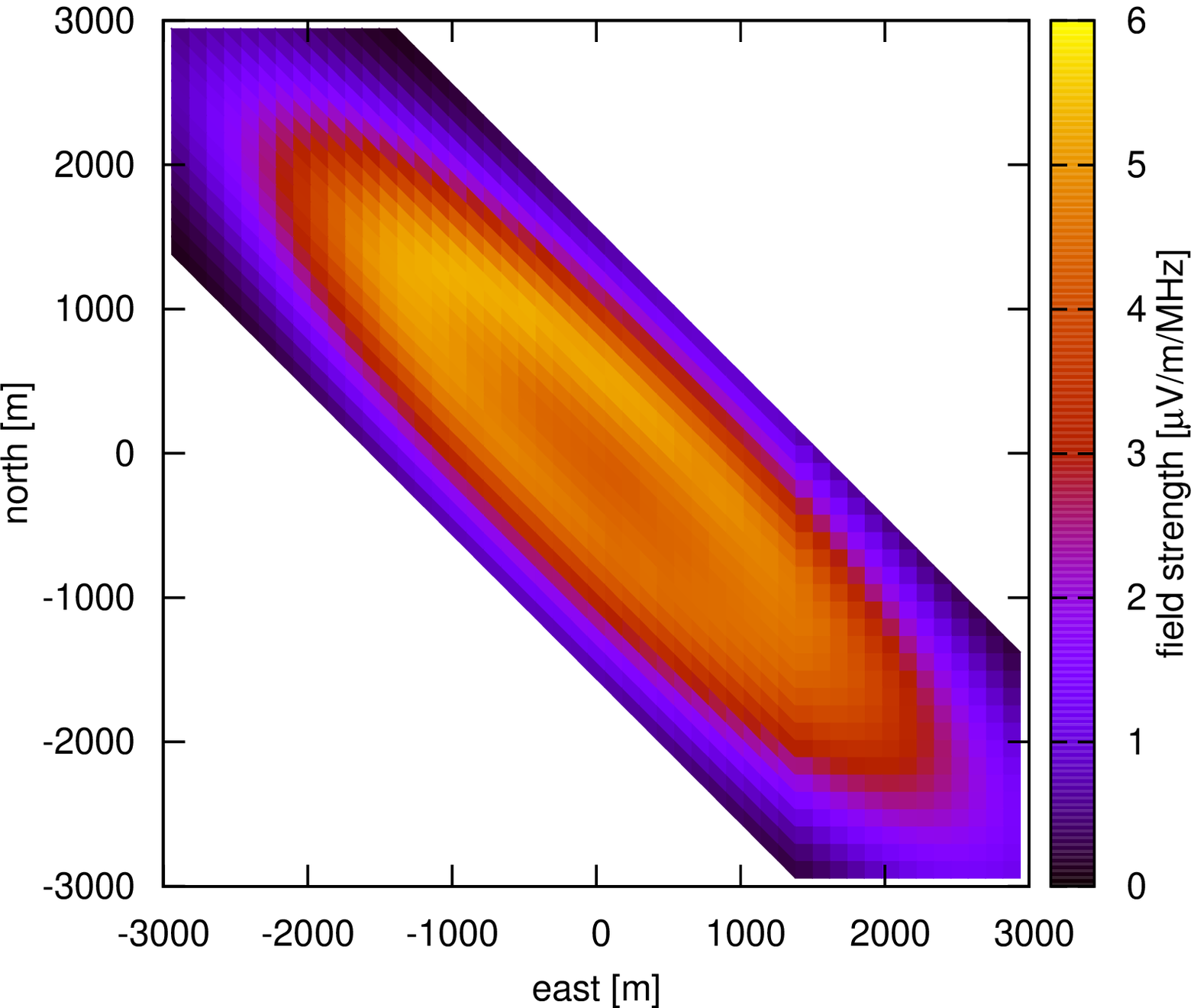} \\
  \includegraphics[width=0.31\textwidth]{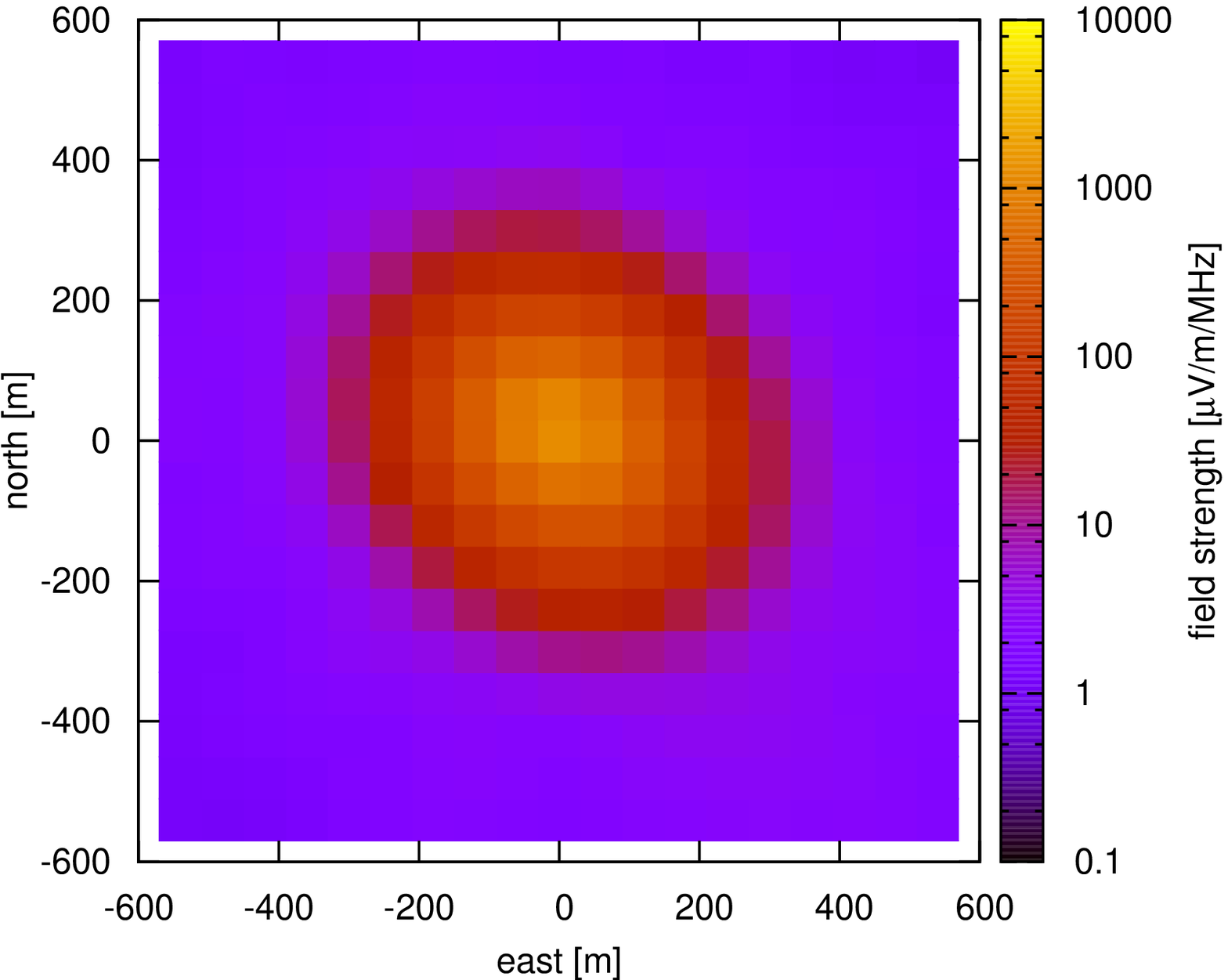} \hspace{0.2cm}
  \includegraphics[width=0.31\textwidth]{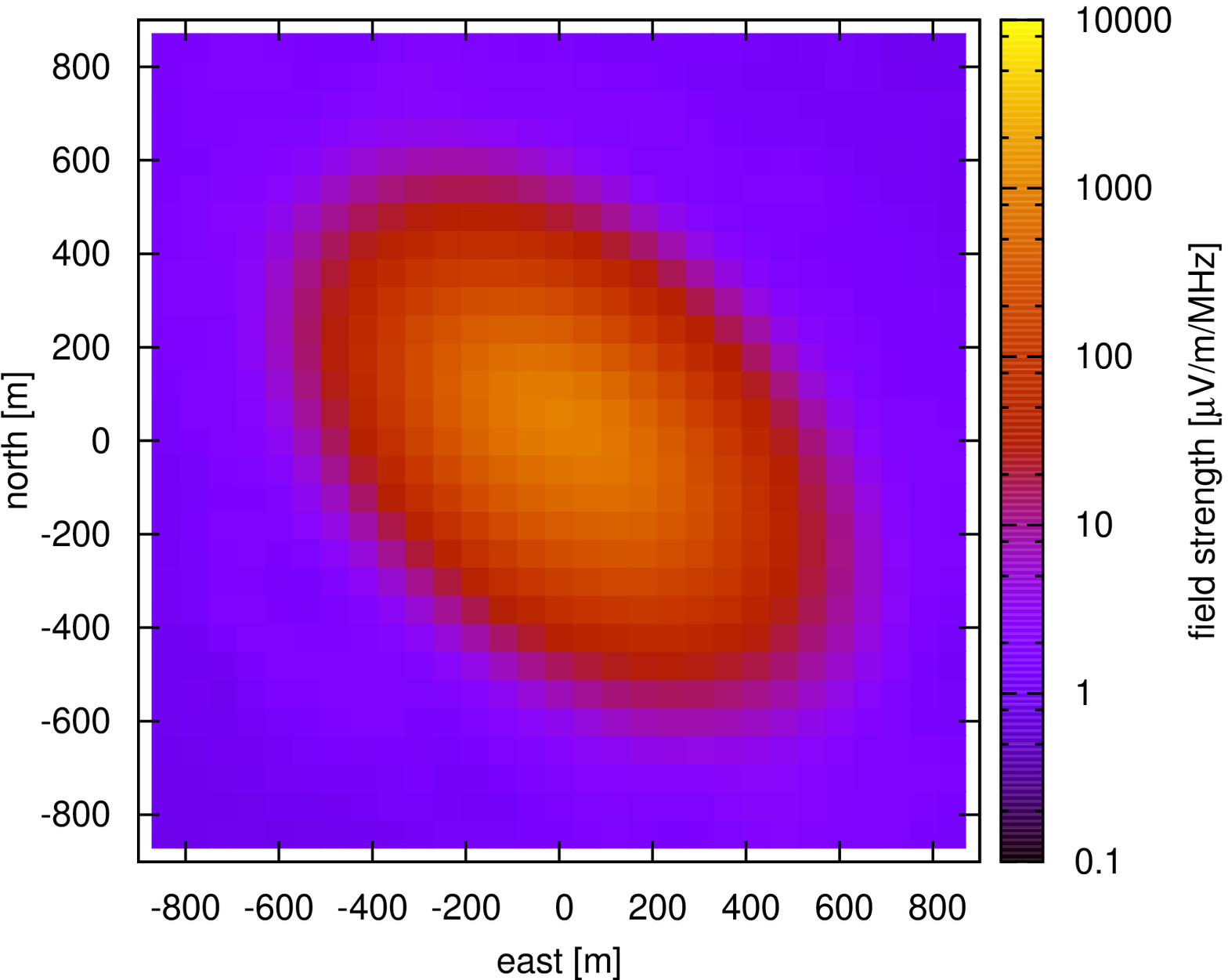} \hspace{0.2cm}
  \includegraphics[width=0.31\textwidth]{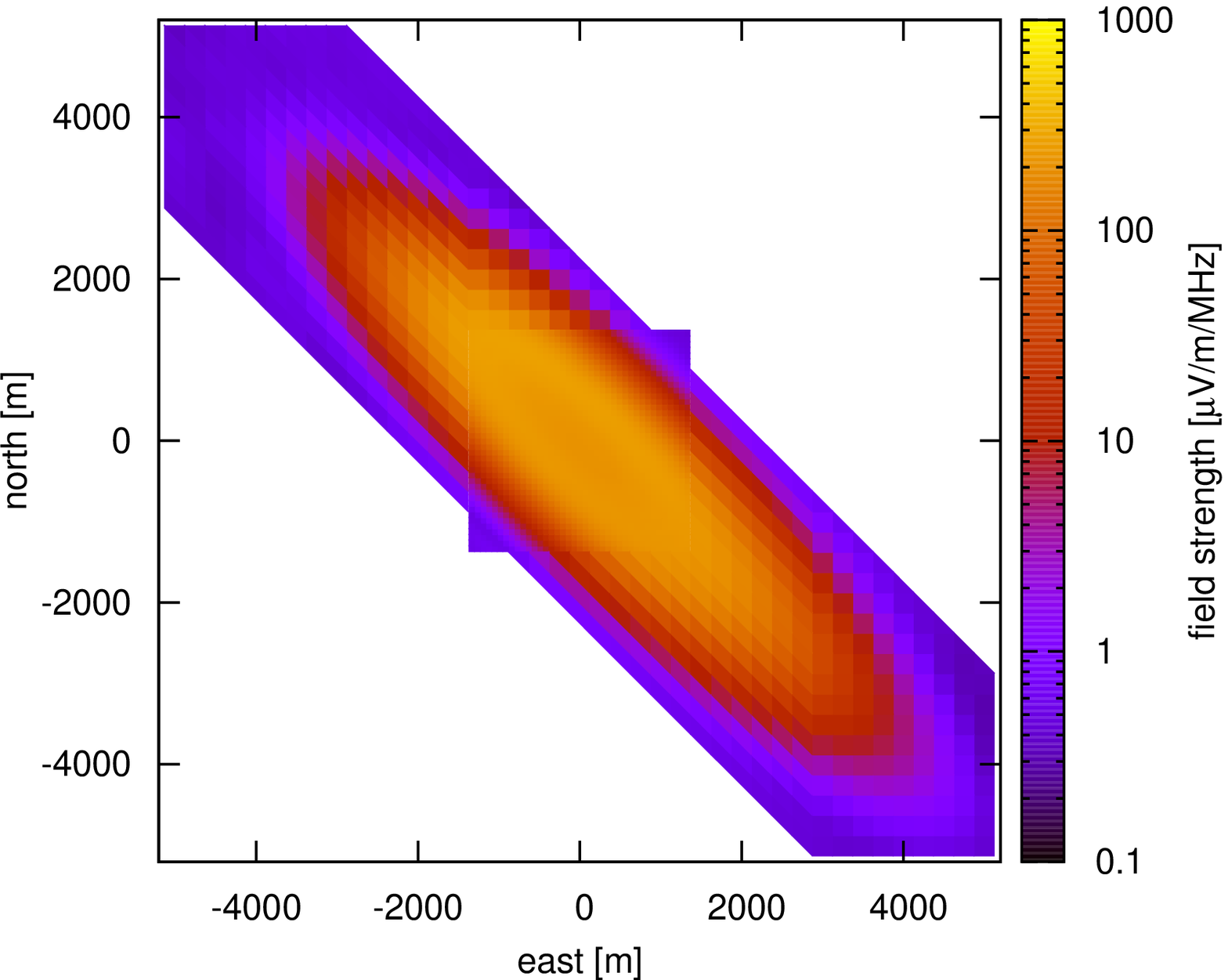}
  \caption{Total spectral electric field amplitude in the observing frequency window 30 to 80 MHz as predicted by CoREAS simulations. The coordinates represent distance on the ground. From left to right: zenith angles of 30$^{\circ}$, 50$^{\circ}$ and 75$^{\circ}$. Top: primary protons with 10$^{18}$ eV, bottom: primary protons with 10$^{19.5}$~eV.}
  \label{fig:mapsinclined}
\end{figure*}

The second set of simulations we present were made for the site of the Auger Engineering Radio Array (AERA \cite{SchroederIcrc2013AERA}), taking into account the observer altitude of 1400~m, the local geomagnetic field (with a strength of 23 nT and an inclination of -38$^{\circ}$) and the observing frequency window of 30 to 80~MHz used by AERA. The same hadronic interaction models as detailed in the previous chapter were used to simulate the radio emission on a dense rectangular grid with up to 2500 observer positions. A thinning level of 10$^{-6}$ with optimized weight limitation was used for the simulations presented here. In figure \ref{fig:mapsinclined}, we show the simulated emission for proton-induced air showers with energies of $10^{18}$~eV (top) and $10^{19.5}$~eV (bottom). From left to right, we demonstrate the changes induced by the increase of the zenith angle of the extensive air showers.

The left-most showers have a zenith angle of $30^{\circ}$. The emission footprint on the ground is mostly circular and falls off very rapidly. At $10^{18}$~eV, the detection threshold of $\approx$ 2 $\mu$V/m/MHz is already reached at radial distances of only 100 to 200 metres. This is consistent with the experience gathered from LOPES \cite{FalckeNature2005}, CODALEMA \cite{ArdouinBelletoileCharrier2005} and AERA \cite{SchroederIcrc2013AERA}. At $10^{19.5}$~eV, we need to keep in mind that due to the applied particle thinning, there is a ``noise floor'' in the simulations so that the spectral amplitudes do not fall below a certain value even as the lateral distance increases. A less aggressive thinning in the simulations would solve this problem. However, it is easily possible to extrapolate from the higher field strengths and conclude that the detection threshold even for $10^{19.5}$~eV is predicted to be reached after at most 400 to 500~m lateral distance for a zenith angle of $30^{\circ}$.

The middle showers have a zenith angle of $50^{\circ}$. The emission footprint becomes markably elliptical and less steep, with lower maximum spectral amplitudes than for the more vertical geometry. The flatter lateral distribution of the radio signal is caused by the greater geometrical distance of the shower particles from the observing radio antennas as $X_{\mathrm{max}}$ (geometrically) recedes with increasing zenith angle \cite{HuegeFalcke2005b,GoussetRavelRoy2004}. The region with a detectable signal grows, however it still remains of the order of only 600~m radius at $10^{18}$~eV and 900~m radius at $10^{19.5}$~eV.

Finally, the right-most showers were simulated with a zenith angle of $75^{\circ}$. The change with respect to $50^{\circ}$ is dramatic. Even for $10^{18}$~eV, CoREAS predicts a detectable amplitude up to distances of 3000~m. For $10^{19.5}$~eV, a detectable signal should still be present at lateral distances of up to 5000~m along the ellipse and roughly 1500~m perpendicular to the elongation of the ellipse. Going to even higher zenith angles will increase the illuminated area even further. The Cherenkov ring visible in the footprint could potentially be exploited to determine $X_{\mathrm{max}}$ for a measured shower.\\

\section{Conclusion}

With CoREAS, a powerful Monte Carlo simulation code for the simulation of radio emission from extensive air showers is openly available. It is being used already by many experiments such as LOPES, LOFAR, AERA, Tunka-REX, ANITA and CROME, and is freely available upon request from the authors.

We have demonstrated the capabilities of CoREAS with simulations for an unthinned vertical air shower, studying the evolution of the emission pattern with increasing frequency. Cherenkov effects lead to a time-compression of the Askaryan and geomagnetic emission, making the air shower radio signal observable at high frequencies for observers sitting on the Cherenkov angle. At GHz frequencies the emission characteristics change noticably, with a clover-leaf pattern appearing which might be related to the similar patterns observed in early ``geosynchrotron emisson'' calculations.

CoREAS is also a very powerful tool for simulation studies aiming at optimizing the design of future radio-detection experiments. We have demonstrated that the footprint of the radio signal at 30 to 80~MHz stays small even at high primary particle energies for showers with zenith angles of up to $50^{\circ}$. However, for zenith angles of $75^{\circ}$ or higher, the footprint becomes extremely large, which means that a sparse radio detector array should readily be able to detect radio emission from air showers at very inclined geometries. In addition, particle detectors mostly measure the muonic component of very inclined air showers, and radio detectors give a pure measurement of the electromagnetic component, making their combination a potentially very powerful tool for cosmic-ray composition studies in particular for very inclined air showers.


\vspace*{0.5cm}
\footnotesize{{\bf Acknowledgment:}{This work has been supported by grant number VH-NG-413 of the Helmholtz Association. The authors would like to thank Ewa Holt for preparing the amplitude maps for inclined air showers and all colleagues who have contributed to fruitful discussions, in particular the Karlsruhe colleagues involved in LOPES, AERA and CROME.}}












\end{document}